\documentclass[fleqn,usenatbib]{mnras}
\pdfoutput=1
\usepackage[T1]{fontenc}
\usepackage{ae,aecompl}
\usepackage{graphicx}
\usepackage{times}
\usepackage{ulem}

\newcommand{\Nbody}{$N$-body }
\newcommand{\MMi}{M/M_{\rm i}}
\newcommand{\rj}{r_{\rm J}}
\newcommand{\rhrj}{r_{\rm h}/r_{\rm J}}
\newcommand{\rtrj}{r_{\rm t}/r_{\rm J}}
\newcommand{\trh}{t_{\rm rh,i}}
\newcommand{\Rh}{R_{\rm h}}

\title[Internal kinematics of star clusters]{Kinematical evolution of tidally limited star clusters: the role of retrograde stellar orbits}

\author[M. Tiongco, E. Vesperini and A.~L. Varri]{
Maria A. Tiongco,$^{1}$\thanks{E-mail: mtiongco@indiana.edu}
Enrico Vesperini,$^{1}$
and Anna Lisa Varri$^{2}$
\\
$^{1}$Department of Astronomy, Indiana University, Bloomington, IN 47405, USA\\
$^{2}$Institute for Astronomy, University of Edinburgh, Royal Observatory, Blackford Hill, Edinburgh EH9 3HJ, UK
}

\date{Accepted XXX. Received YYY; in original form ZZZ}

\pubyear{2016}

\begin{document}
\label{firstpage}
\pagerange{\pageref{firstpage}--\pageref{lastpage}}
\maketitle

\begin{abstract}

The presence of an external tidal field often induces significant dynamical evolutionary effects on the internal kinematics of star clusters. Previous studies investigating the restricted three-body problem with applications to star cluster dynamics have shown that unbound stars on retrograde orbits (with respect to the direction of the cluster's orbit) are more stable against escape than prograde orbits, and predicted that a star cluster might acquire retrograde rotation through preferential escape of stars on prograde orbits. In this study we present evidence of this prediction, but we also illustrate that there are additional effects that cannot be accounted for by the preferential escape of prograde orbits alone. Specifically, in the early evolution, initially underfilling models increase their fraction of retrograde stars without losing significant mass, and acquire a retrograde angular velocity. We attribute this effect to the development of preferentially eccentric/radial orbits in the outer regions of star clusters as they are expanding into their tidal limitation.

We explore the implications of the evolution of the fraction of prograde and retrograde stars for the evolution of the cluster internal rotation, and its dependence on the initial structural properties. Although all the systems studied here evolve towards an approximately solid-body internal rotation with angular velocity equal to about half of the angular velocity of the cluster orbital motion around the host galaxy, the evolutionary history of the radial profile of the cluster internal angular velocity depends on the cluster initial structure. 

\end{abstract}

\begin{keywords}
galaxies: star clusters: general; Galaxy: globular clusters: general; methods: numerical; 
\end{keywords}

\section{Introduction}

The combined effects of internal two-body relaxation and the external potential of the host galaxy play a major role in determining the structural and kinematical evolution of globular clusters, the rate at which clusters lose mass and their lifetimes.  Early works in this regard include, for example,  \citet{vesperini1997}, \citet{giersz1997}, and \citet{baumgardt2003}, and a number of more recent investigations including \citet{madrid2012}, \citet{webb2014}, \citet{haghi2014}, and \citet{zonoozi2016} have further explored these issues.
The complex interplay between internal and external evolutionary processes alters in a non-trivial way the initial cluster properties and complicates any effort aimed at understanding which of the current cluster properties were imprinted at the epoch of its formation and which are instead the result of its evolutionary history \citep[see e.g.][for some studies attempting to reconstruct the cluster initial properties]{zonoozi2011,zonoozi2014,webb2015}.

The observational characterization of the internal kinematics of several Galactic globular clusters will soon reach an unprecedented level of richness, thanks the synergy between the forthcoming astrometric data provided by {\it Gaia} and {\it HST} \citep[see e.g.][]{richer2013,bellini2014,bellini2015,watkins2015,sollima2015}, and a number of ESO/VLT spectroscopic programs \citep[see e.g.][]{lanzoni2013,lapenna2015,lardo2015}.  
The knowledge of cluster internal kinematical properties opens the way to major advances in our understanding of cluster formation and dynamical evolution and of the role played by different internal and external evolutionary processes in shaping the current cluster properties.

In \citet{tiongco2016}, we have presented the results of a survey of simulations aimed at exploring the link between anisotropy in the velocity distribution and cluster initial structural properties, dynamical phase and mass loss history.

In this paper, we will focus on the characterization of the stellar orbital properties of a star cluster evolving in the external potential of the host galaxy, and we will discuss their implications on the global kinematics of the cluster.

A number of numerical studies \citep{keenan1975,fukushige2000,ernst2007,zotos2015} have shown that unbound stars still residing within the cluster are more stable against escape if they are on retrograde orbits (with respect to the direction of the cluster's orbit), with non-trivial implications on the definition of boundary of a tidally perturbed stellar system in coordinate and velocity space \citep[see][]{read2006,gajda2015}. 
As pointed out by \citet{keenan1975}, the preferential loss of prograde stars and the ensuing excess of retrograde stars implies that the process of mass loss may be accompanied by the development of an internal retrograde rotation. A study of the evolution of the velocity moments of \Nbody models performed by \citet{baumgardt2003} has provided some evidence of the presence of such a behaviour, especially in the outer parts of tidally perturbed star clusters.

At the same time, the interaction with the external tidal field of the host galaxy might in part offset this effect and drive the system towards an internal rotation synchronous with that associated with the cluster orbital motion. Such a condition of ``synchronization'' of the internal and orbital motion (``tidal locking'') is indeed a standard assumption underlying many investigations of the dynamics of star clusters, which often rely on the classical formulation of the Hill's problem \citep[see e.g.][]{chandrasekhar1942,heggie2003}.

In this paper we present a detailed analysis of the long-term evolution of the fractions of the prograde and retrograde orbiting populations in \Nbody models of star clusters evolving in an external tidal field. We study the evolution of clusters models for a range of different initial structural properties, and we explore how these properties affect the evolution of the fraction of prograde and retrograde stars and the implications for the cluster internal kinematics, especially its angular momentum. We explore in detail the dynamical processes behind the evolution of the fraction of prograde and retrograde stars, both for bound stars and for the population of unbound stars still residing within the cluster (``potential escapers'').

The outline of the paper is the following. In Section 2, we describe our method and the initial conditions adopted. We present our results in Section 3, and in Section 4 we summarize our conclusions.

\section{Method and Initial Conditions}
\label{sec:method}

We have carried out a survey of \Nbody simulations using the {\sevensize NBODY6} code \citep{aarseth2003} accelerated by a GPU \citep{nitadori2012}; the simulations were run on the {\sevensize BIG RED II} cluster at Indiana University.

For all the simulations, we have assumed clusters on circular orbits in the host galaxy external field modelled as that of a point-mass.  The coordinate system in which equations of motion are solved is corotating with the cluster around the host galaxy with angular velocity $\Omega$ \citep[see e.g.][]{heggie2003}. For the primary set of simulations, we assume that the system has no internal rotation in this reference frame; in a non-rotating reference frame, this corresponds to a rotation synchronous  with that of the cluster around the host galaxy. Hereafter we shall refer to these set of models as `locked' models. 
  To explore the potential dependence of the evolution of the kinematics on whether or not the system is initially locked, we have also considered some models that are initially non-rotating in a non-rotating frame reference.
  In the rotating frame of reference described above these models have a retrograde solid-body rotation with angular velocity equal to $-\Omega$.  We will refer to these models as `unlocked' models, and identify them by adding U to the model ID (see Table \ref{tab:details}).

  We consider systems with $N=$ 16 384 equal-mass particles, and particles moving beyond a distance from the cluster centre equal to two times the Jacobi radius, $\rj$, are removed from the simulation.  All the simulations are run until 75\% of the initial cluster mass is lost.

The initial conditions of our systems are \Nbody realizations of isotropic, \citet{king1966} models, with dimensionless central potential $W_0=7$,
and different values of the initial filling factor, defined as the ratio of the three-dimensional half-mass radius to the Jacobi radius $\rhrj$ or equivalently, the ratio of the model truncation radius to the Jacobi radius $\rtrj$ (see Table \ref{tab:details}). These models were originally presented in the context of a study of the long-term evolution of the velocity anisotropy of tidally limited star clusters  and we refer the reader to \citet{tiongco2016} for a complete description (notice that there was a typo in reporting the $(\rhrj)_{\rm i}$ of the model KF0125 in Table 1 of \citealt{tiongco2016}.  The correct value is shown in this paper).  

By setting the value of $\rj$ according to the desired filling factor, $\Omega$ is determined as

\begin{equation}
\Omega = \sqrt{\frac{GM}{3\rj^3}}.
\end{equation}

\noindent As $\rj$ increases for more underfilling models, $\Omega$ decreases and this implies  that the initial conditions and evolution of the unlocked versions of our underfilling models do not differ significantly from those of the corresponding locked versions. 
In consideration of this, we focus our attention only on unlocked models with relatively large filling factors, KF1U, KF075U and KF05U.

We emphasize that in the rest of the paper, whenever we discuss cluster rotation or refer to the cluster internal angular velocity, $\omega$, we will refer to the angular velocity {\it as measured in the co-rotating reference frame described above}. In this reference frame, prograde and retrograde orbits are, respectively, those with $\omega$ larger or smaller than zero and an orbit with zero angular velocity is that of a star in corotation with the cluster around the host galaxy.
\begin{table}
\caption{Summary of simulations}
\label{tab:details}
\begin{tabular}{@{}lcc}
\hline
Model ID & 
$(\rhrj)_{\rm i}$ & 
$(\rtrj)_{\rm i}$\\
\hline
KF1 & 0.116 & 1.0\\
KF075 & 0.087 & 0.75\\
KF05 & 0.058 & 0.5\\
KF025 & 0.029 & 0.25\\
KF0125 & 0.015 & 0.125\\
\\
KF1U & 0.116 & 1.0\\
KF075U & 0.087 & 0.75\\
KF05U & 0.058 & 0.5\\
\hline
\end{tabular}
\end{table}

\section{Results}

\subsection{Evolution of the angular velocity profile}

\begin{figure*}
	\includegraphics[width=6.6in,height=6.6in]{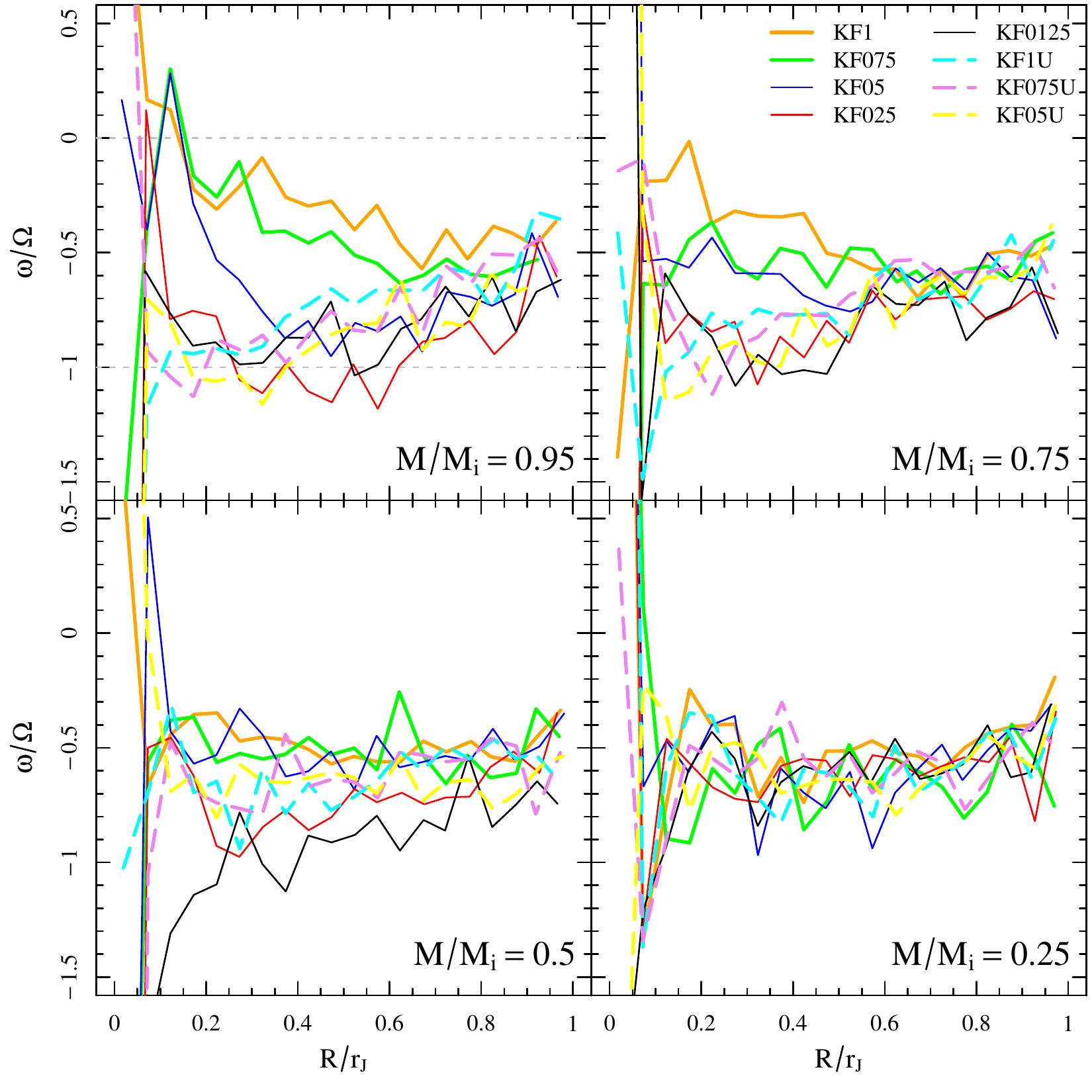}
    \caption{Angular velocity profiles of all models at the respective times when they have fractions of mass remaining $\MMi$ = 0.95, 0.75, 0.5, and 0.25 (here $M$ is defined as the total mass enclosed in the Jacobi radius, $\rj$, i.e., including both bound and unbound stars).  Also, shown in the first panel (top left) as dashed lines are the starting values for the locked ($\omega/\Omega$ = 0) and unlocked models ($\omega/\Omega$ = -1).}
    \label{fig:angvel}
\end{figure*}

We approach the characterization of the kinematical properties of the models by studying the evolution of the radial profile of the internal angular velocity, $\omega$.

The angular velocity is measured in the direction parallel to the $z$- axis, binning the particles in  equally thick cylindrical shells with heights encompassing the entire cluster at different projected radii, $R$.  We have combined 5 snapshots around the desired times.  In the construction of the radial profiles, all particles within the Jacobi radius have been used.

The radial profiles of $\omega$, normalized to the angular velocity of the cluster's orbit around the host galaxy, $\Omega$, are illustrated in Fig. \ref{fig:angvel}. Each panel offers a comparison between the profiles of all models evaluated at four different stages of their evolution, identified by the amount of mass that is retained by the system ($\MMi=0.95, 0.75, 0.50, 0.25$), where $M$ denotes the total mass of all particles enclosed in the Jacobi radius.

All the models that are initially locked
start with $\omega_{\rm i}(R)=0$, while
the unlocked models have, in the rotating frame, a solid body retrograde rotation ($\omega_{\rm i}(R)=-\Omega$).  In all cases, as the
locked systems start to lose mass they all acquire a retrograde
rotation in their outer regions (beyond approximately 0.1 $\rj$; see the last panel of Fig.~\ref{fig:angvel}).

Besides providing a quantitative measurement of the effect of the preferential loss of stars in prograde orbits on the cluster kinematics, this figure reveals several interesting and non-trivial results.

First of all, it appears  
that the early evolution of the radial profile of $\omega$ depends on the cluster initial filling factor for the locked models. 
Strongly underfilling models develop a strong retrograde angular
velocity already at the beginning of their evolution (see the behaviour of KF0125 and KF025 in the first panel), while more filling models (i.e., KF075, KF1) are initially characterized by a more moderate retrograde rotation.  As the evolution proceeds and the systems lose more mass, the behaviour of the angular velocity of the filling models
converge to an approximately retrograde solid-body rotation such that $\omega/\Omega\simeq-0.5$, 
starting with the outermost regions and later extending to the inner
regions.  The model KF05 is an intermediate case, having a stronger
retrograde angular velocity than the more filling models at $\MMi$ =
0.95, but follows a similar behaviour by the time it has
lost 50\% of its mass.  The most underfilling models, KF025 and
KF0125, and the unlocked models take longer to converge but eventually
at $\MMi$ = 0.25 all models share the same approximately solid-body rotation with
$\omega\simeq$-0.5 $\Omega$ at $R/r_J>0.1$.

Both initially locked and unlocked systems evolve towards a common radial profile of $\omega$ but the relative importance of different dynamical processes in driving evolution towards the final solid-body rotation profile depends on the cluster initial properties. The more filling and initially locked systems lose their initial synchronous rotation as a result of the preferential loss of prograde stars (in this respect, the role of the so-called ``potential escapers'' will be discussed in detail in Section \ref{sec:bound}). Initially underfilling models, on the other hand, experience an early evolutionary phase during which their outer regions develop 
a strong retrograde rotation which may be interpreted in light of an additional mechanism not relying on mass loss (we will further discuss in Sections \ref{sec:globalnntot} and \ref{sec:plateau} the dynamics underpinning this mechanism).

In the subsequent evolution, while the interaction with the tidal field drives the systems towards a synchronization of the internal angular velocity with that of the cluster orbital motion around the host galaxy ($\omega/\Omega=0$ in the co-rotating reference frame), the preferential loss of particles on prograde orbits counteracts in part this effect. The net result is the retrograde approximate solid-body rotation shown in Fig. \ref{fig:angvel}.

We wish to emphasize that for the purpose of the current investigation, we will focus exclusively on understanding the physical mechanisms that are responsible for the presence of an approximate solid-body rotation in the intermediate to outer parts of the models under consideration. The role played by the presence of non-vanishing total angular momentum in the dynamical evolution of collisional stellar systems is indeed a much broader topic \citep[see e.g.][]{goodman1983,einsel1999}, and in particular, the long-term kinematical evolution of models with differential rotation in a tidal field has an added layer of complexity that has been only partially explored \citep[see e.g.][]{ernst2007,hong2013}. We plan to fully address the interplay between internal rotation and external tidal field in a future study (Tiongco et al., in prep.).

\subsection{Early evolution of the global fraction of retrograde orbits}
\label{sec:globalnntot}

\begin{figure}
	\includegraphics[width=3.3in,height=3.3in]{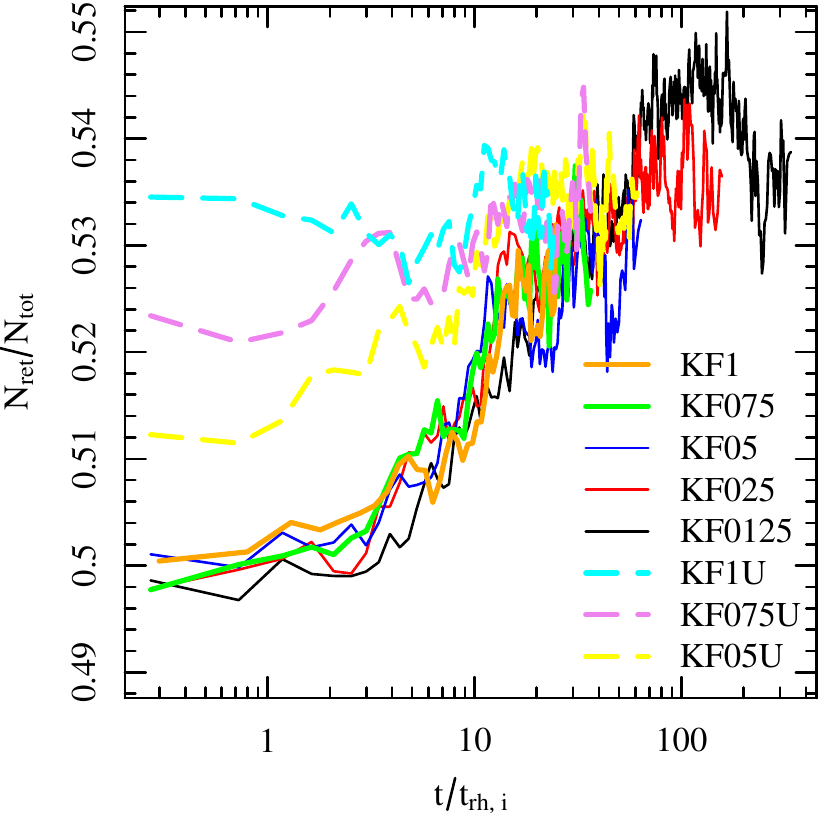}
    \caption{The instantaneous global fraction of retrograde stars as a function of time normalized to the initial half-mass relaxation time (all particles within the Jacobi radius have been considered).}
    \label{fig:nntot}
\end{figure}

\begin{figure}
	\includegraphics[width=3.3in,height=3.3in]{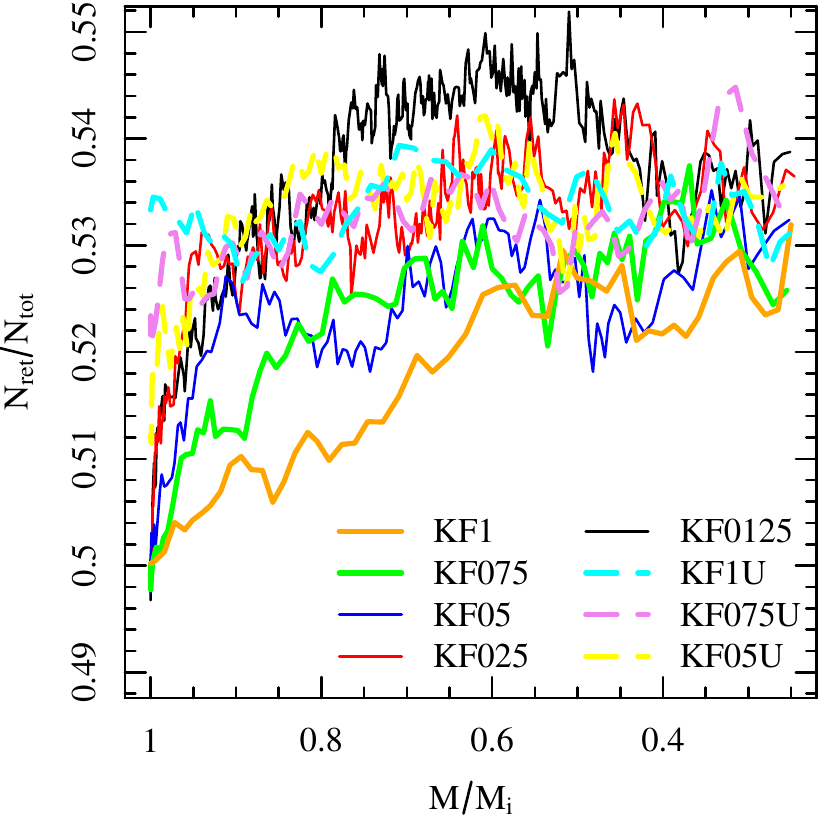}
    \caption{Same as Fig. \ref{fig:nntot} but as a function of mass remaining in the system ($M$ is defined as in Fig.~\ref{fig:angvel}).}
    \label{fig:nntot2}
\end{figure}

\begin{figure}
	\includegraphics[width=3.3in]{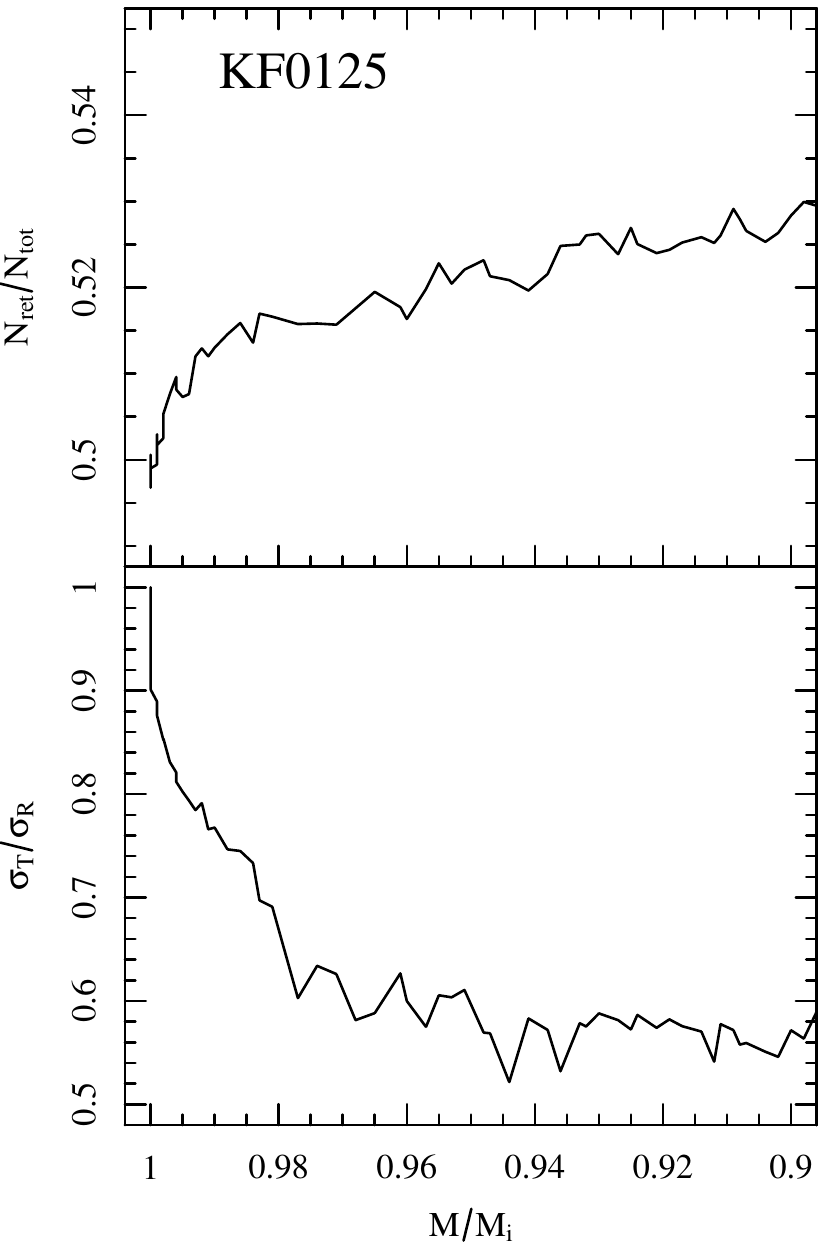}
    \caption{Top panel: A zoomed-in version of $N_{\rm ret}/N_{\rm tot}$ vs mass remaining for model KF0125.  Bottom panel: Evolution of the velocity anisotropy for the outer regions of this model ($R > 3\Rh$, where $\Rh$ is the 2D projected half-mass radius)}.
    \label{fig:massani}
\end{figure}

\begin{figure*}
	\includegraphics[width=6.6in,height=6.6in]{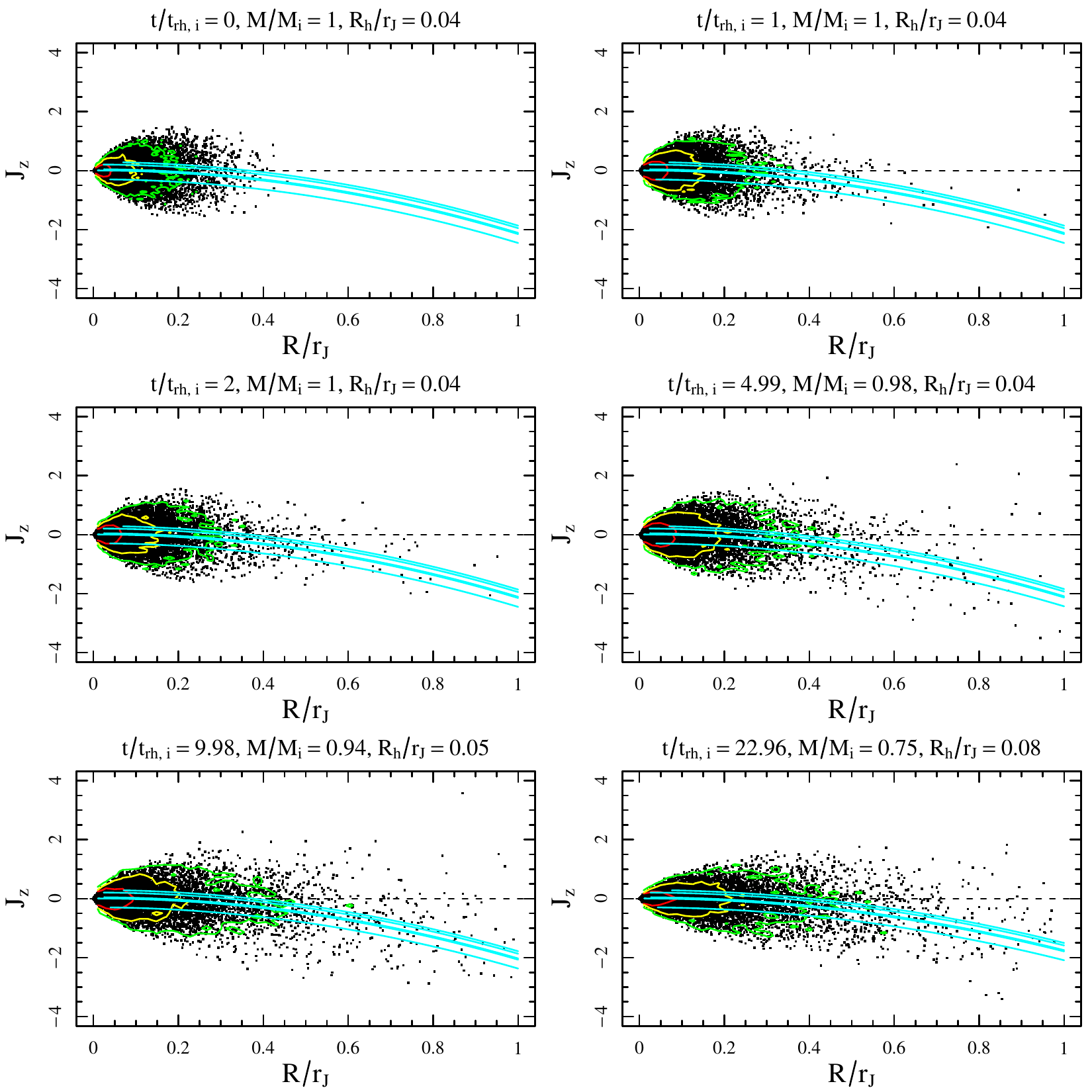}
    \caption{$J_z$ vs. $R$ normalized to the Jacobi radius for several snapshots for model KF05.  The contours give an indication of the density of the points, with red being the high density to green being low density, and the curved lines show the predicted trajectories of a few particles due to conservation of angular momentum only.  The total mass $M$ is defined as in Fig.~\ref{fig:angvel}.}
    \label{fig:jzvsr}
\end{figure*}

The possible development of retrograde rotation has been often associated exclusively with the preferential loss of prograde orbiting stars. In this section, we delve further into this issue and present the results of our analysis aimed at quantifying the evolution of the populations of prograde and retrograde orbiting stars.  In Fig. \ref{fig:nntot}, we show the time evolution of $N_{\rm ret}/N_{\rm tot}$, the global fraction of the total number of stars within the Jacobi radius on retrograde orbits.

We focus our attention first on the initially locked
models; the time evolution of the global fraction of stars on retrograde orbits is characterized by an initial  progressive growth, at a rate approximately independent of the initial filling factor of the system  followed by a phase in which the fraction of retrograde orbits reaches a {\it plateau} and remains approximately constant and equal to about 0.52-0.54 until the end of the simulations. The fraction of retrograde orbits at the {\it plateau} is slightly larger and reached later in more underfilling systems (see e.g. in Fig. \ref{fig:nntot} the slight differences in evolution of the fraction of retrograde orbits in the KF05, KF025, and KF0125 models).

On the other hand, the unlocked models begin their evolution with a 
larger fraction of retrograde orbiting particles, 
and the time evolution of the total fraction of retrograde orbits is less 
significant, especially for the more filling models (see model KF1U in Fig. \ref{fig:nntot}).

When the evolution of the global fraction of retrograde orbits is represented as a function of the fraction of the initial mass remaining in the system (see Fig. \ref{fig:nntot2}), some interesting aspects of the dynamics behind the evolution of the fraction of prograde and retrograde stars emerge. Initially locked, tidally filling models require a significant mass loss in order to show an appreciable growth of the fraction of retrograde orbits (see models KF1 and KF075, denoted with orange and green lines in Fig.~\ref{fig:nntot2}), which therefore may be interpreted as associated with the loss of stars on prograde orbits.

But we emphasize here that in initially locked {\it underfilling} models, the initial evolution of the fraction of prograde and retrograde orbits is {\it not} associated with mass loss. This is particularly evident for the KF0125 and the KF025 models (depicted with black and red lines in Fig.~\ref{fig:nntot2}), and indicates that for these systems there must be an additional internal mechanism affecting the fraction of prograde and retrograde stars before the preferential escape of prograde stars becomes important. We will now provide evidence that such a mechanism is indeed tied to the development of a halo of stars on eccentric/radial orbits driven by the  effects of two-body relaxation.

As discussed in \citet{tiongco2016} and in a number of previous studies \citep[see e.g.][]{giersz1994,giersz1997,giersz2011} initially isotropic star clusters that are either isolated or underfilling, develop during their long-term evolution an outer halo characterized by a radially anisotropic velocity distribution.

As a particle moves outwards on a highly eccentric orbit, angular momentum conservation (in a non-rotating frame of reference) implies that its rotational velocity around the z-axis slows down, and therefore the particle's rotational velocity tends to become retrograde with respect to the rotating frame.

By inspecting a zoomed-in version of the global fraction of retrograde particles in, for example, the most underfilling system KF0125 (see Fig. \ref{fig:massani}, top panel), the growth of the fraction of retrograde particles is particularly evident, in the absence of significant mass loss. When compared to the growth of the radial velocity anisotropy in the outer regions (> 3$\Rh$, where $\Rh$ is the 2D projected half-mass radius; see Fig.~\ref{fig:massani}, bottom panel), the timescale of the increase appears to be consistent.  
To measure the degree of anisotropy we use the ratio of the projected tangential to the radial velocity dispersion, $\sigma_{\rm T}/\sigma_{\rm R}$, measured on the cluster orbital plane.

We note that a similar interpretation applies also to unlocked systems, in which the difference between the behaviour of filling and underfilling models is less prominent (especially because, as discussed above, their initial fraction of retrograde orbits is initially higher compared to the corresponding locked systems), but still appreciable.

To provide a more quantitative characterization of the dynamics of the stars populating the outer regions of the star clusters, we explore the evolution of the models in the plane $(R,J_z)$, where $J_z$ is the $z$ component of the angular momentum calculated in the rotating frame of reference and $R$ is projected distance from the cluster centre measured on the $(x,y)$ plane (the analysis performed on model KF05 is illustrated in Fig.~ \ref{fig:jzvsr}). As particles start to populate the halo, they do so mostly with retrograde velocities and the outer regions are dominated by retrograde particles. A few trajectories on this plane determined by assuming angular momentum conservation are also plotted to further illustrate this effect.

\subsection{The radial variation of the fraction of prograde and retrograde orbits and the origin of the {\it plateau}}
\label{sec:plateau}

\begin{figure}
	\includegraphics[width=3.3in,height=3.3in]{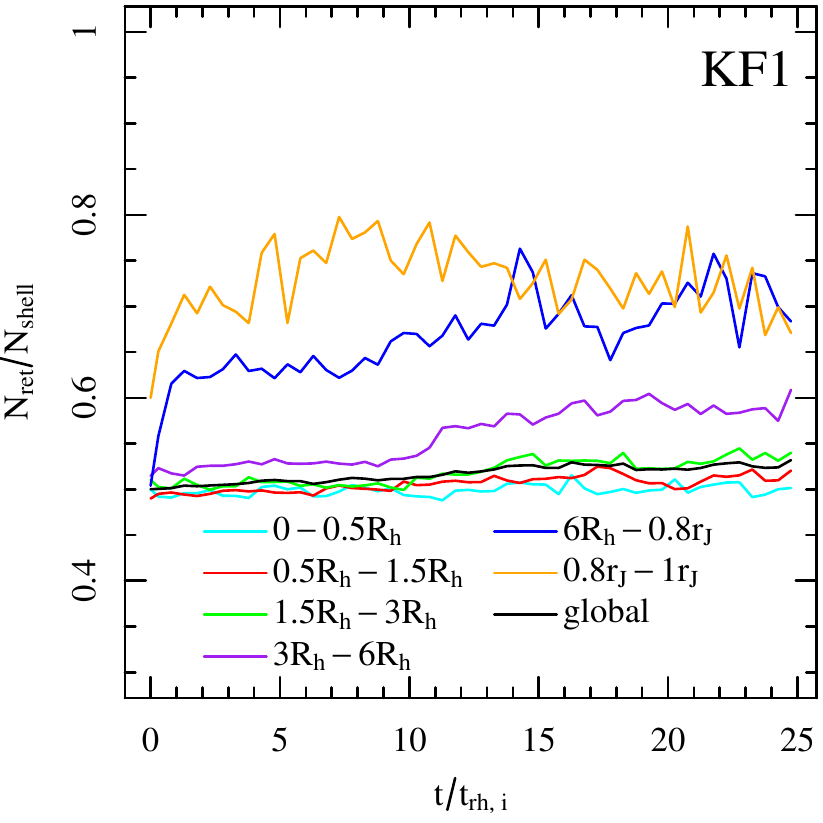}
    \caption{The time evolution of model KF1 of the fraction of retrograde stars within several radial cylindrical shells, with respect to the current number of stars within the respective shell.  Note that the black line corresponds to the orange line for KF1 in Fig.~\ref{fig:nntot}.}
    \label{fig:nntot_KF1}
\end{figure}

\begin{figure}
	\includegraphics[width=3.3in,height=3.3in]{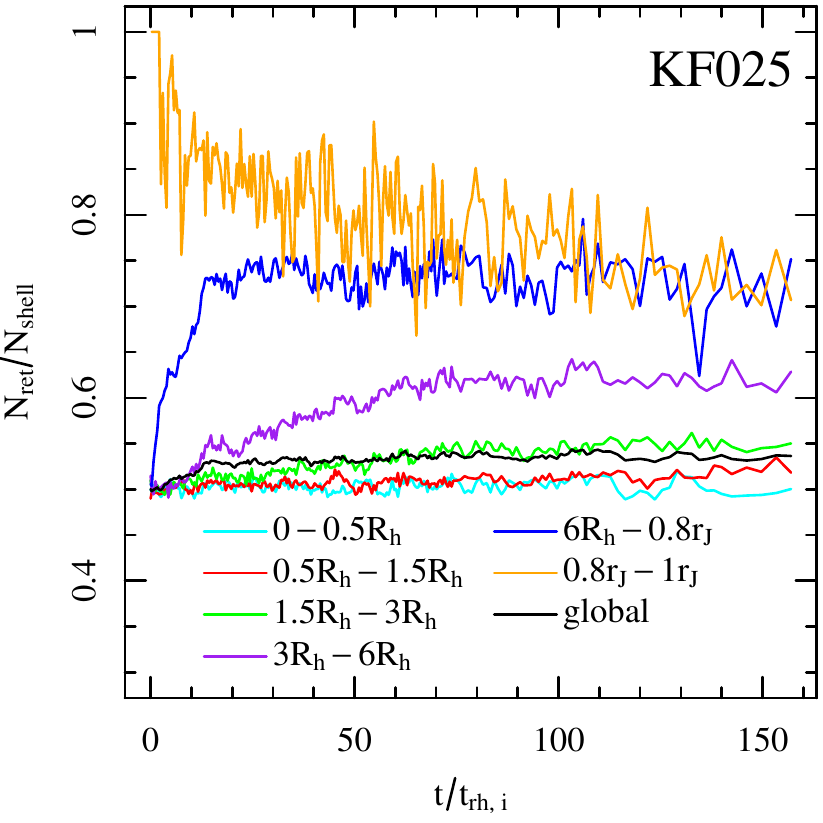}
    \caption{Similar to Fig. \ref{fig:nntot_KF1} for model KF025.  Note that the black line corresponds to the red line for KF025 in Fig.~\ref{fig:nntot}.}
    \label{fig:nntot_KF025}
\end{figure}

While in the previous section we have explored the evolution of the fraction of retrograde orbits mostly from a global perspective, we now wish to investigate how the orbital distribution evolves within different regions of a tidally perturbed star cluster.

We first consider
the filling system KF1 (see Fig.~\ref{fig:nntot_KF1}), by exploring the evolution of the fraction of retrograde stars within several radial shells, again using cylindrical shells with heights encompassing the entire cluster (notice that the fraction of retrograde particles for each shell is calculated with respect to the number of stars within each shell and not the total number of stars within the system).  Indeed, the outermost shells show the strongest signature of asymmetry between the prograde and retrograde populations while in the innermost shells ($R<1.5 \Rh$), the fraction of retrograde particles remains approximately constant and equal to its initial value. The growth of the fraction retrograde orbits in the intermediate to outer shells is characterized by an early growth followed by a late flattening similar to that already seen in Fig. \ref{fig:nntot} for the evolution of the global fraction of retrograde orbits.  For this model, the evolution of the fraction of retrograde particles is therefore dominated by the preferential loss of prograde particles although, as we will further discuss in Section \ref{sec:bound}, the additional mechanism tied to the development of radial anisotropy is affecting also this model.

As for star clusters in the underfilling regime, we perform the same analysis on the representative model KF025 (see Fig.~\ref{fig:nntot_KF025}).
In this case the outermost regions near the cluster Jacobi radius are initially empty, and, as illustrated in Fig.~\ref{fig:nntot2}, the fraction of retrograde particles starts to increase well before any significant mass loss. As discussed in Section \ref{sec:globalnntot}, we ascribe this initial increase in the fraction of retrograde particles to the development of a halo of particles on highly-eccentric/radial orbits. Fig. \ref{fig:nntot_KF025}  further illustrates how retrograde particles largely dominate the cluster outermost regions. Afterwards, when the system starts to lose significant mass, the preferential loss of prograde stars becomes important.

This exploration of the radial dependence of the fraction of prograde and retrograde orbits allows us also to provide an interpretation of the origin of the {\it plateau} observed in the second phase of the evolution of the global fraction of retrograde stars (see Figs.~\ref{fig:nntot} and \ref{fig:nntot2}). 
We suggest that the origin of such plateau lies in the balancing act between the mass loss due to the expansion of the outer layers across the tidal boundary and the continued ``production'' of retrograde orbits through the mechanism associated with the ejection of stars from the core to the halo on highly eccentric/radial orbits (a process further enhanced in the post-core-collapse).

This interpretation seems to be well supported by the behaviour of the intermediate to outer shells of our models, which show a moderate increase of the fraction of retrograde orbits, followed by a stabilization or even a mild decline, especially for the last shell of initially underfilled systems (see purple, blue, and orange lines in Figs. \ref{fig:nntot_KF1} and \ref{fig:nntot_KF025}).
Additional evidence of the role of the tidal stripping of a cluster outer layers in the evolution of cluster anisotropy and structural properties has been discussed by \citet{giersz1997}.

\subsection{Fraction of prograde and retrograde orbits for bound and unbound particles}

\label{sec:bound}

As studied in detail by \citet{fukushige2000}, an unbound particle still within a cluster Jacobi radius can take many dynamical times before actually escaping the cluster. This can lead to the formation of a population of potential escapers residing within the cluster and affecting the cluster structural and kinematical properties \citep[][]{kuepper2010}.
In this section we further analyse our results in the context of energetically bound or unbound particles. In the frame co-rotating with a star cluster on a circular orbit in a point-mass galaxy, the energy of a star is given by \citep[see e.g.][]{fukushige2000}

\begin{equation}
E = \frac{v^2}{2} + \phi_{\rm c} + \frac{1}{2}\Omega^2(z^2-3x^2),
\end{equation}

\noindent where $v$ is the speed of the star and $\phi_{\rm c}$ is the potential of the cluster.  Following \citet{fukushige2000} we define the critical energy as the potential at the Lagrangian points

\begin{equation}
E_{\rm crit} = -\frac{3GM}{2\rj},
\end{equation}

\noindent and identify  bound or unbound stars as those having an energy, respectively, below or above, $E_{\rm crit}$.  

\begin{figure}
	\includegraphics[width=3.3in]{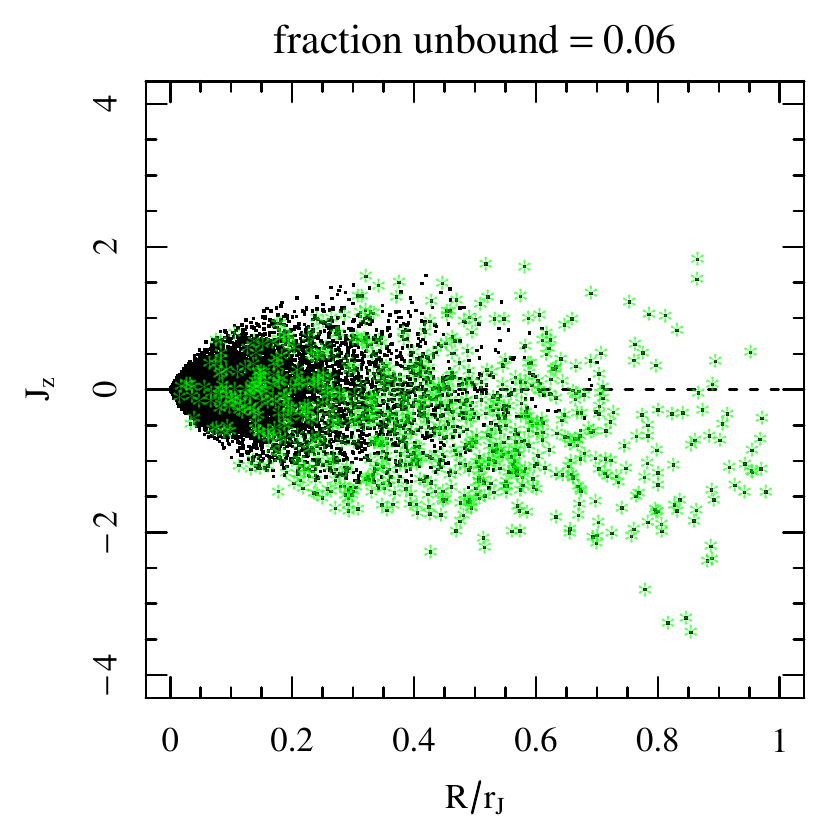}
    \caption{The same as the last panel of Fig. \ref{fig:jzvsr}, but with the unbound particles highlighted in green.}
    \label{fig:jzvr_u}
\end{figure}

We find that potential escapers may dominate the outermost regions of clusters. For example, Fig. \ref{fig:jzvr_u} shows a snapshot of the plane $(J_z,R)$ for model KF05, with the potential escapers highlighted in green.  It is evident that almost all of the outer cluster stars beyond approximately 0.5 $\rj$ are potential escapers.  This result is in general agreement with the findings of \citet{kuepper2010}, who first showed that the behaviour of the surface brightness and velocity dispersion radial profiles are significantly affected by the presence of an outer population of potential escapers.

\begin{figure}
	\includegraphics[width=3.3in]{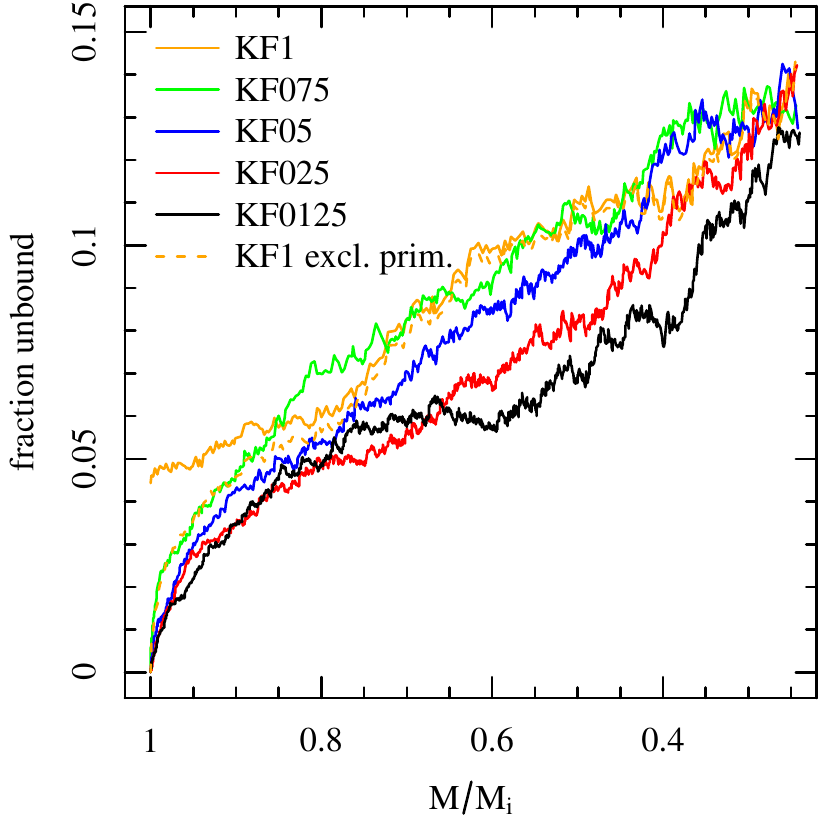}
    \caption{The fraction of unbound particles as a function of fraction of mass remaining in the cluster for all locked models.  For KF1, there are two separate lines for including primordially unbound particles and excluding them.}
    \label{fig:unbd_mass}
\end{figure}

The fraction of unbound particles as a function of the fraction of the mass remaining in the cluster for all our locked models is illustrated in Fig. \ref{fig:unbd_mass}.
As discussed in \citet{fukushige2000} and \citet{baumgardt2001}, initial conditions set by sampling, for example, a King model may include a fraction of stars which are unbound when an external tidal field is added. This is the case for our KF1 model. For this model we show  the evolution of the fraction of potential escapers both with and without  this population of primordial potential escapers.
All the other models do not have any primordial potential escapers. 

Fig. \ref{fig:unbd_mass} also shows that the fraction of potential escapers  increases significantly over time reaching a fraction of the remaining mass equal to about 15\% by the end of our simulations. Such a behaviour is qualitatively consistent with the findings reported by \citet{baumgardt2001}, in which tidally perturbed models (initialised as critically filling King model $W_0=3$, i.e. characterized by a much higher fraction of primordial potential escapers, about 15 per cent) possess approximately 20 per cent of potential escapers during the final stages of evolution. 

\begin{figure}
	\includegraphics[width=3.3in,height=5in]{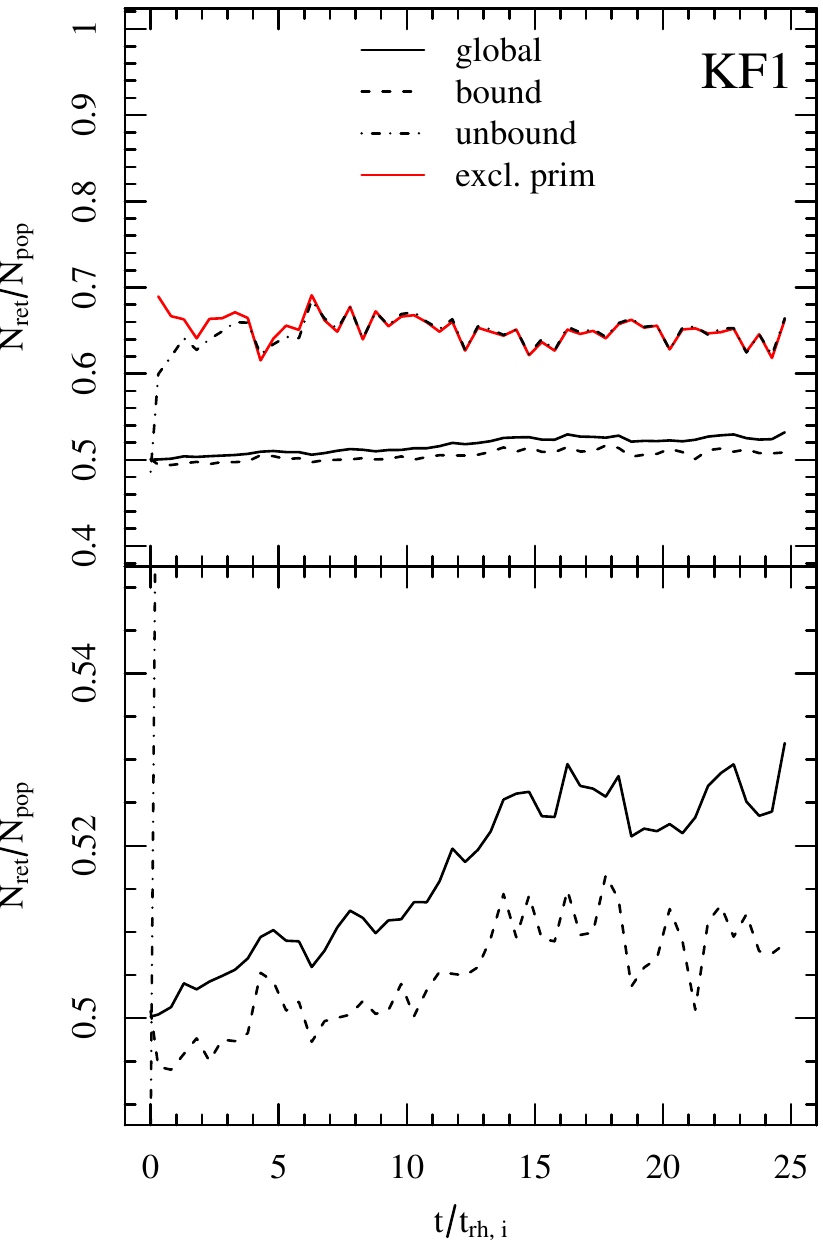}
    \caption{The time evolution of model KF1 of the fraction of retrograde stars within the entire system, and within energetically bound and unbound particles.  The bottom panel is a zoomed in version.  The red line shows the fraction of unbound particles that do not include the primordially unbound stars. Note that the  black solid lines in both panels correspond to the orange line for KF1 in Fig.~\ref{fig:nntot}.}
    \label{fig:nntot_KF1en}
\end{figure}

\begin{figure}
	\includegraphics[width=3.3in,height=5in]{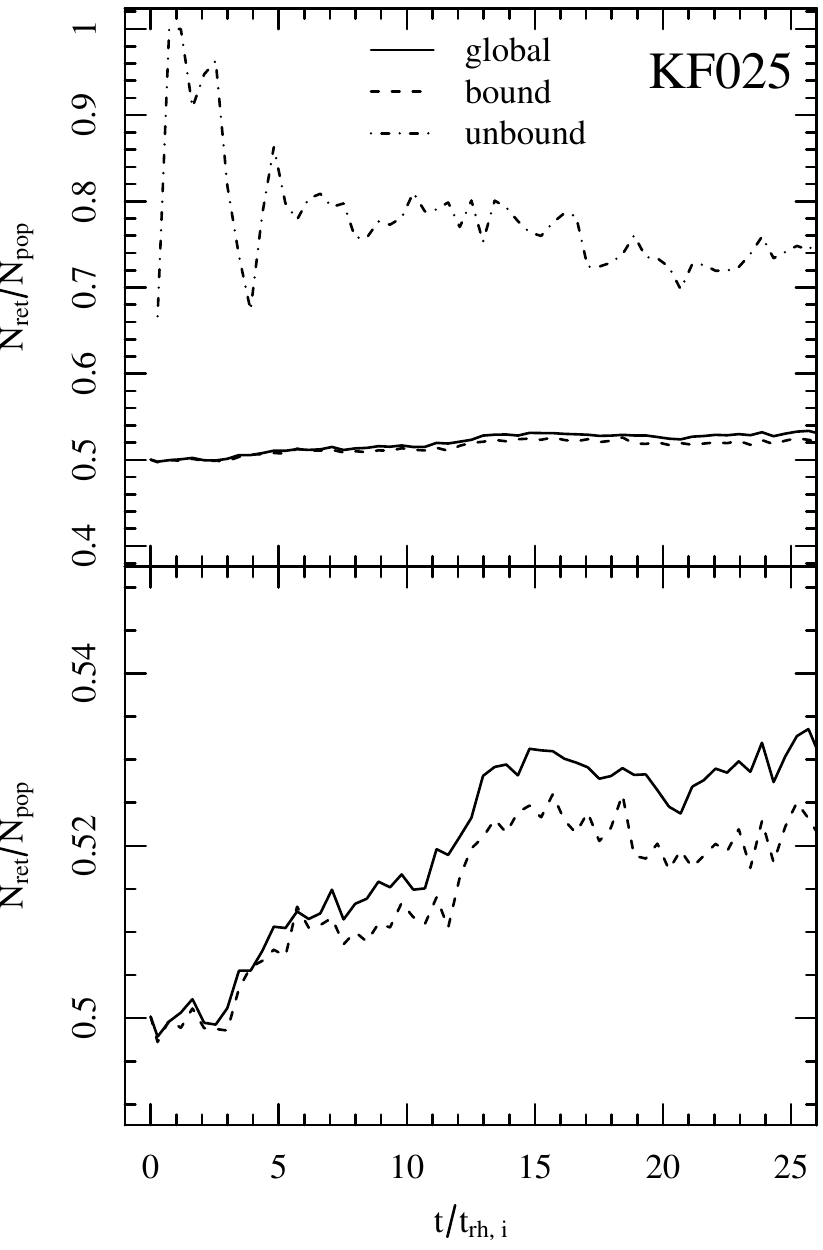}
    \caption{Same as Fig. \ref{fig:nntot_KF1en} for model KF025. Note that the black solid lines in both panels correspond to the red line for KF025 in Fig.~\ref{fig:nntot}.}
    \label{fig:nntot_KF025en}
\end{figure}

We have examined the evolution of the fraction of prograde and retrograde particles within the populations of bound and unbound particles in models KF1 and KF025, as representative of the two filling regimes (see, Figs. \ref{fig:nntot_KF1en} and \ref{fig:nntot_KF025en}, respectively). 
The larger fraction of retrograde particles in the population of unbound particles is the result of the two effects discussed in Section \ref{sec:globalnntot}: the preferential escape of prograde particles and the fact that, in the transition from bound to unbound, some particles move on more eccentric/radial orbits becoming retrograde as they move outwards.

It is interesting that also the bound population shows an asymmetry in the fractions of prograde and retrograde particles.  The growth in the fraction of retrograde particles in this case must be due to particles that develop more eccentric orbits while remaining bound to the cluster. The development of an excess of radially-biased orbits (as measured by the presence of radial anisotropy, see bottom panel of Fig. \ref{fig:massani}) 
in the bound population is therefore the cause of the increase in the fraction of retrograde particles in this population.
As discussed in \citet{tiongco2016}, even the most filling  King model (KF1) was characterized after core collapse by the development of radial anisotropy (although very weak) in the inner regions near the half-mass radius (see Fig. 10 in \citealt{tiongco2016}) and indeed, Fig. \ref{fig:nntot_KF1en} shows that for the KF1 model the slight increase in the fraction of retrograde particles starts only after core collapse at $t>10-12 \trh$.

\section{Conclusions}
We have studied the effects of the host galaxy tidal field and star escape on the long-term evolution of a star cluster's orbital distribution and internal kinematics. In particular, we have focused our attention on the evolution in the number of stars on prograde and retrograde orbits with respect to the cluster's orbit about the galaxy and the implications for the evolution of the cluster internal rotation.  We have carried out a number of \Nbody simulations using isotropic \citet{king1966} models varying the initial filling factor, and experimenting with models without rotation in the standard frame of reference co-rotating with an angular velocity equal to that of the cluster around the host galaxy, i.e. ``locked", as well as with ``unlocked'' systems.

Previous studies investigating the restricted three-body problem of a star orbiting in the potential of the cluster and that of the cluster host galaxy have shown that unbound stars on retrograde orbits are more stable against escape, and predicted that a cluster might acquire retrograde rotation through preferential escape of stars on prograde orbits \citep{keenan1975}.  In this study, we present evidence of this prediction in action and quantify its effect: the rotation curve of all our models evolves towards a retrograde rotational velocity beginning with the outer regions and moving inward, and the cluster's fraction of retrograde orbiting particles slightly increases over time.

However, we have shown there are additional effects that cannot be accounted for by preferential escape of prograde orbits alone.  In the early evolution, initially underfilling models increase their fraction of retrograde stars without losing significant mass and acquire a retrograde angular velocity.  We attribute this effect to the development of preferentially eccentric/radial orbits in the outer regions of star clusters 
as 
they are expanding into their tidal limitation.   
Such an excess of elongated orbits can be measured in terms of the presence of a radially-biased anisotropy in the velocity space \citep[see e.g.][]{giersz1994,giersz1997,giersz2011,tiongco2016,zocchi2016}.
As a particle moves outwards on a highly eccentric orbit, angular momentum conservation (in a non-rotating
frame of reference) implies that its rotational velocity around the $z$-axis slows down and therefore that the particle rotational velocity tends to become retrograde with respect to the rotating frame. We have provided evidence that particles on these eccentric orbits populate the cluster halo mainly on retrograde orbits.  We have also explored the time evolution of the radial variation of the fraction of retrograde particles and shown that, as the cluster evolves, the fraction of retrograde orbits increases with the distance from the cluster centre and the outer regions are increasingly populated by stars on retrograde orbits.

It is important to emphasize that once an underfilling cluster has expanded into its tidal limitation and is able to lose particles more rapidly, the preferential escape of prograde orbits becomes important and these models behave similarly to the initially filling models.

For all initially locked models, we observe that the rate at which the fraction of retrograde stars increases slows down until it stops and the fraction remains almost constant and equal to about 0.52-0.54 for the rest of the simulation.  We have interpreted such a plateau as the result of the balance between the mass loss associated with the tidal stripping of the cluster outer layers expanding across the tidal boundary and the continued generation of retrograde orbits through the mechanism associated with the ejection of stars from the core to halo on preferentially highly eccentric/radial orbits.

The behaviour of the unlocked models considered in this study may be interpreted in terms of the same mechanisms illustrated above, although their evolution start with slightly larger fractions of retrograde orbits, which eventually converge to values similar to those reached in the simulations of non-rotating systems.

The interplay between the evolution of the fraction of prograde and retrograde stars and the effects of the tidal field leads to a rotational velocity radial profile with an angular velocity approximately constant and equal to about half of the angular velocity of the cluster orbital motion around the centre of the host galaxy. Although all the systems studied eventually reach this profile after losing a large fraction of their initial mass, the evolutionary history of the angular velocity radial profile depends on the cluster initial structural and kinematical properties.

Finally, we have explored the orbital properties of energetically bound and unbound particles. The fraction of unbound particles within the cluster (potential escapers) increases with time and by the end of our simulations (when the systems have lost 75\% of their initial mass) they comprise almost 15\% of the remaining mass \citep[see also][]{baumgardt2001,kuepper2010} and  dominate the outer regions beyond approximately 0.5 $\rj$.  Unbound particles have a higher fraction of retrograde particles when compared to the fraction calculated for bound particles or the whole system.  Interestingly, also within the population of bound particles the retrograde fraction slightly increases with time; we ascribe this evolution to the increase of elongated orbits (i.e., an increase of radial anisotropy) in the bound population.

In the future we will investigate the behaviour of models with higher degree of rotation, and thus beginning with a stronger difference in the fraction of prograde and retrograde orbits.

Our investigation therefore further illustrates the complex effects of internal and external dynamical processes on the cluster phase-space density distribution.  As observations from ESO/VLT, {\it HST} and the {\it Gaia} mission shed light on the internal kinematical properties of star clusters, theoretical efforts aimed at building a framework for the interpretation of these results must include the development of new families of dynamical models including features produced by dynamical processes such as  different fractions of prograde/retrograde orbits and the presence of a significant fraction of unbound potential escapers (see Daniel, Heggie \& Varri in prep).
\bibliographystyle{mnras}
\bibliography{references} 

\begin{thebibliography}{}
\makeatletter
\relax
\def\mn@urlcharsother{\let\do\@makeother \do\$\do\&\do\#\do\^\do\_\do\%\do\~}
\def\mn@doi{\begingroup\mn@urlcharsother \@ifnextchar [ {\mn@doi@}
  {\mn@doi@[]}}
\def\mn@doi@[#1]#2{\def\@tempa{#1}\ifx\@tempa\@empty \href
  {http://dx.doi.org/#2} {doi:#2}\else \href {http://dx.doi.org/#2} {#1}\fi
  \endgroup}
\def\mn@eprint#1#2{\mn@eprint@#1:#2::\@nil}
\def\mn@eprint@arXiv#1{\href {http://arxiv.org/abs/#1} {{\tt arXiv:#1}}}
\def\mn@eprint@dblp#1{\href {http://dblp.uni-trier.de/rec/bibtex/#1.xml}
  {dblp:#1}}
\def\mn@eprint@#1:#2:#3:#4\@nil{\def\@tempa {#1}\def\@tempb {#2}\def\@tempc
  {#3}\ifx \@tempc \@empty \let \@tempc \@tempb \let \@tempb \@tempa \fi \ifx
  \@tempb \@empty \def\@tempb {arXiv}\fi \@ifundefined
  {mn@eprint@\@tempb}{\@tempb:\@tempc}{\expandafter \expandafter \csname
  mn@eprint@\@tempb\endcsname \expandafter{\@tempc}}}

\bibitem[\protect\citeauthoryear{{Aarseth}}{{Aarseth}}{2003}]{aarseth2003}
{Aarseth} S.~J.,  2003, {Gravitational N-Body Simulations}.
Cambridge Univ. Press, Cambridge

\bibitem[\protect\citeauthoryear{{Baumgardt}}{{Baumgardt}}{2001}]{baumgardt2001}
{Baumgardt} H.,  2001, \mn@doi [\mnras] {10.1046/j.1365-8711.2001.04272.x},
  \href {http://adsabs.harvard.edu/abs/2001MNRAS.325.1323B} {325, 1323}

\bibitem[\protect\citeauthoryear{{Baumgardt} \& {Makino}}{{Baumgardt} \&
  {Makino}}{2003}]{baumgardt2003}
{Baumgardt} H.,  {Makino} J.,  2003, \mn@doi [\mnras]
  {10.1046/j.1365-8711.2003.06286.x}, \href
  {http://adsabs.harvard.edu/abs/2003MNRAS.340..227B} {340, 227}

\bibitem[\protect\citeauthoryear{{Bellini} et~al.}{{Bellini}
  et~al.}{2014}]{bellini2014}
{Bellini} A.,  et~al., 2014, \mn@doi [\apj] {10.1088/0004-637X/797/2/115},
  \href {http://adsabs.harvard.edu/abs/2014ApJ...797..115B} {797, 115}

\bibitem[\protect\citeauthoryear{{Bellini} et~al.}{{Bellini}
  et~al.}{2015}]{bellini2015}
{Bellini} A.,  et~al., 2015, \mn@doi [\apjl] {10.1088/2041-8205/810/1/L13},
  \href {http://adsabs.harvard.edu/abs/2015ApJ...810L..13B} {810, L13}

\bibitem[\protect\citeauthoryear{{Chandrasekhar}}{{Chandrasekhar}}{1942}]{chandrasekhar1942}
{Chandrasekhar} S.,  1942, {Principles of stellar dynamics}.
Univ. Chicago Press, Chicago, IL

\bibitem[\protect\citeauthoryear{{Einsel} \& {Spurzem}}{{Einsel} \&
  {Spurzem}}{1999}]{einsel1999}
{Einsel} C.,  {Spurzem} R.,  1999, \mn@doi [\mnras]
  {10.1046/j.1365-8711.1999.02083.x}, \href
  {http://adsabs.harvard.edu/abs/1999MNRAS.302...81E} {302, 81}

\bibitem[\protect\citeauthoryear{{Ernst}, {Glaschke}, {Fiestas}, {Just}  \&
  {Spurzem}}{{Ernst} et~al.}{2007}]{ernst2007}
{Ernst} A.,  {Glaschke} P.,  {Fiestas} J.,  {Just} A.,   {Spurzem} R.,  2007,
  \mn@doi [\mnras] {10.1111/j.1365-2966.2007.11602.x}, \href
  {http://adsabs.harvard.edu/abs/2007MNRAS.377..465E} {377, 465}

\bibitem[\protect\citeauthoryear{{Fukushige} \& {Heggie}}{{Fukushige} \&
  {Heggie}}{2000}]{fukushige2000}
{Fukushige} T.,  {Heggie} D.~C.,  2000, \mn@doi [\mnras]
  {10.1046/j.1365-8711.2000.03811.x}, \href
  {http://adsabs.harvard.edu/abs/2000MNRAS.318..753F} {318, 753}

\bibitem[\protect\citeauthoryear{{Gajda} \& {{\L}okas}}{{Gajda} \&
  {{\L}okas}}{2016}]{gajda2015}
{Gajda} G.,  {{\L}okas} E.~L.,  2016, \mn@doi [\apj]
  {10.3847/0004-637X/819/1/20}, \href
  {http://adsabs.harvard.edu/abs/2016ApJ...819...20G} {819, 20}

\bibitem[\protect\citeauthoryear{{Giersz} \& {Heggie}}{{Giersz} \&
  {Heggie}}{1994}]{giersz1994}
{Giersz} M.,  {Heggie} D.~C.,  1994, \mn@doi [\mnras]
  {10.1093/mnras/268.1.257}, \href
  {http://adsabs.harvard.edu/abs/1994MNRAS.268..257G} {268, 257}

\bibitem[\protect\citeauthoryear{{Giersz} \& {Heggie}}{{Giersz} \&
  {Heggie}}{1997}]{giersz1997}
{Giersz} M.,  {Heggie} D.~C.,  1997, \mn@doi [\mnras]
  {10.1093/mnras/286.3.709}, \href
  {http://adsabs.harvard.edu/abs/1997MNRAS.286..709G} {286, 709}

\bibitem[\protect\citeauthoryear{{Giersz} \& {Heggie}}{{Giersz} \&
  {Heggie}}{2011}]{giersz2011}
{Giersz} M.,  {Heggie} D.~C.,  2011, \mn@doi [\mnras]
  {10.1111/j.1365-2966.2010.17648.x}, \href
  {http://adsabs.harvard.edu/abs/2011MNRAS.410.2698G} {410, 2698}

\bibitem[\protect\citeauthoryear{{Goodman}}{{Goodman}}{1983}]{goodman1983}
{Goodman} J.~J.,  1983, PhD thesis, Princeton Univ., NJ.

\bibitem[\protect\citeauthoryear{{Haghi}, {Hoseini-Rad}, {Zonoozi}  \&
  {K{\"u}pper}}{{Haghi} et~al.}{2014}]{haghi2014}
{Haghi} H.,  {Hoseini-Rad} S.~M.,  {Zonoozi} A.~H.,   {K{\"u}pper} A.~H.~W.,
  2014, \mn@doi [\mnras] {10.1093/mnras/stu1714}, \href
  {http://adsabs.harvard.edu/abs/2014MNRAS.444.3699H} {444, 3699}

\bibitem[\protect\citeauthoryear{{Heggie} \& {Hut}}{{Heggie} \&
  {Hut}}{2003}]{heggie2003}
{Heggie} D.,  {Hut} P.,  2003, {The Gravitational Million-Body Problem: A
  Multidisciplinary Approach to Star Cluster Dynamics}.
Cambridge Univ. Press, Cambridge

\bibitem[\protect\citeauthoryear{{Hong}, {Kim}, {Lee}  \& {Spurzem}}{{Hong}
  et~al.}{2013}]{hong2013}
{Hong} J.,  {Kim} E.,  {Lee} H.~M.,   {Spurzem} R.,  2013, \mn@doi [\mnras]
  {10.1093/mnras/stt099}, \href
  {http://adsabs.harvard.edu/abs/2013MNRAS.430.2960H} {430, 2960}

\bibitem[\protect\citeauthoryear{{Keenan} \& {Innanen}}{{Keenan} \&
  {Innanen}}{1975}]{keenan1975}
{Keenan} D.~W.,  {Innanen} K.~A.,  1975, \mn@doi [\aj] {10.1086/111744}, \href
  {http://adsabs.harvard.edu/abs/1975AJ.....80..290K} {80, 290}

\bibitem[\protect\citeauthoryear{{King}}{{King}}{1966}]{king1966}
{King} I.~R.,  1966, \mn@doi [\aj] {10.1086/109857}, \href
  {http://adsabs.harvard.edu/abs/1966AJ.....71...64K} {71, 64}

\bibitem[\protect\citeauthoryear{{K{\"u}pper}, {Kroupa}, {Baumgardt}  \&
  {Heggie}}{{K{\"u}pper} et~al.}{2010}]{kuepper2010}
{K{\"u}pper} A.~H.~W.,  {Kroupa} P.,  {Baumgardt} H.,   {Heggie} D.~C.,  2010,
  \mn@doi [\mnras] {10.1111/j.1365-2966.2010.17084.x}, \href
  {http://adsabs.harvard.edu/abs/2010MNRAS.407.2241K} {407, 2241}

\bibitem[\protect\citeauthoryear{{Lanzoni} et~al.}{{Lanzoni}
  et~al.}{2013}]{lanzoni2013}
{Lanzoni} B.,  et~al., 2013, \mn@doi [\apj] {10.1088/0004-637X/769/2/107},
  \href {http://adsabs.harvard.edu/abs/2013ApJ...769..107L} {769, 107}

\bibitem[\protect\citeauthoryear{{Lapenna}, {Origlia}, {Mucciarelli},
  {Lanzoni}, {Ferraro}, {Dalessandro}, {Valenti}  \& {Cirasuolo}}{{Lapenna}
  et~al.}{2015}]{lapenna2015}
{Lapenna} E.,  {Origlia} L.,  {Mucciarelli} A.,  {Lanzoni} B.,  {Ferraro}
  F.~R.,  {Dalessandro} E.,  {Valenti} E.,   {Cirasuolo} M.,  2015, \mn@doi
  [\apj] {10.1088/0004-637X/798/1/23}, \href
  {http://adsabs.harvard.edu/abs/2015ApJ...798...23L} {798, 23}

\bibitem[\protect\citeauthoryear{{Lardo} et~al.}{{Lardo}
  et~al.}{2015}]{lardo2015}
{Lardo} C.,  et~al., 2015, \mn@doi [\aap] {10.1051/0004-6361/201425036}, \href
  {http://adsabs.harvard.edu/abs/2015A%26A...573A.115L} {573, A115}

\bibitem[\protect\citeauthoryear{{Madrid}, {Hurley}  \& {Sippel}}{{Madrid}
  et~al.}{2012}]{madrid2012}
{Madrid} J.~P.,  {Hurley} J.~R.,   {Sippel} A.~C.,  2012, \mn@doi [\apj]
  {10.1088/0004-637X/756/2/167}, \href
  {http://adsabs.harvard.edu/abs/2012ApJ...756..167M} {756, 167}

\bibitem[\protect\citeauthoryear{{Nitadori} \& {Aarseth}}{{Nitadori} \&
  {Aarseth}}{2012}]{nitadori2012}
{Nitadori} K.,  {Aarseth} S.~J.,  2012, \mn@doi [\mnras]
  {10.1111/j.1365-2966.2012.21227.x}, \href
  {http://adsabs.harvard.edu/abs/2012MNRAS.424..545N} {424, 545}

\bibitem[\protect\citeauthoryear{{Read}, {Wilkinson}, {Evans}, {Gilmore}  \&
  {Kleyna}}{{Read} et~al.}{2006}]{read2006}
{Read} J.~I.,  {Wilkinson} M.~I.,  {Evans} N.~W.,  {Gilmore} G.,   {Kleyna}
  J.~T.,  2006, \mn@doi [\mnras] {10.1111/j.1365-2966.2005.09959.x}, \href
  {http://adsabs.harvard.edu/abs/2006MNRAS.367..387R} {367, 387}

\bibitem[\protect\citeauthoryear{{Richer}, {Heyl}, {Anderson}, {Kalirai},
  {Shara}, {Dotter}, {Fahlman}  \& {Rich}}{{Richer} et~al.}{2013}]{richer2013}
{Richer} H.~B.,  {Heyl} J.,  {Anderson} J.,  {Kalirai} J.~S.,  {Shara} M.~M.,
  {Dotter} A.,  {Fahlman} G.~G.,   {Rich} R.~M.,  2013, \mn@doi [\apjl]
  {10.1088/2041-8205/771/1/L15}, \href
  {http://adsabs.harvard.edu/abs/2013ApJ...771L..15R} {771, L15}

\bibitem[\protect\citeauthoryear{{Sollima}, {Baumgardt}, {Zocchi}, {Balbinot},
  {Gieles}, {H{\'e}nault-Brunet}  \& {Varri}}{{Sollima}
  et~al.}{2015}]{sollima2015}
{Sollima} A.,  {Baumgardt} H.,  {Zocchi} A.,  {Balbinot} E.,  {Gieles} M.,
  {H{\'e}nault-Brunet} V.,   {Varri} A.~L.,  2015, \mn@doi [\mnras]
  {10.1093/mnras/stv1079}, \href
  {http://adsabs.harvard.edu/abs/2015MNRAS.451.2185S} {451, 2185}

\bibitem[\protect\citeauthoryear{{Tiongco}, {Vesperini}  \& {Varri}}{{Tiongco}
  et~al.}{2016}]{tiongco2016}
{Tiongco} M.~A.,  {Vesperini} E.,   {Varri} A.~L.,  2016, \mn@doi [\mnras]
  {10.1093/mnras/stv2574}, \href
  {http://adsabs.harvard.edu/abs/2016MNRAS.455.3693T} {455, 3693}

\bibitem[\protect\citeauthoryear{{Vesperini} \& {Heggie}}{{Vesperini} \&
  {Heggie}}{1997}]{vesperini1997}
{Vesperini} E.,  {Heggie} D.~C.,  1997, \mn@doi [\mnras]
  {10.1093/mnras/289.4.898}, \href
  {http://adsabs.harvard.edu/abs/1997MNRAS.289..898V} {289, 898}

\bibitem[\protect\citeauthoryear{{Watkins}, {van der Marel}, {Bellini}  \&
  {Anderson}}{{Watkins} et~al.}{2015}]{watkins2015}
{Watkins} L.~L.,  {van der Marel} R.~P.,  {Bellini} A.,   {Anderson} J.,  2015,
  \mn@doi [\apj] {10.1088/0004-637X/803/1/29}, \href
  {http://adsabs.harvard.edu/abs/2015ApJ...803...29W} {803, 29}

\bibitem[\protect\citeauthoryear{{Webb} \& {Leigh}}{{Webb} \&
  {Leigh}}{2015}]{webb2015}
{Webb} J.~J.,  {Leigh} N.~W.~C.,  2015, \mn@doi [\mnras]
  {10.1093/mnras/stv1780}, \href
  {http://adsabs.harvard.edu/abs/2015MNRAS.453.3278W} {453, 3278}

\bibitem[\protect\citeauthoryear{{Webb}, {Leigh}, {Sills}, {Harris}  \&
  {Hurley}}{{Webb} et~al.}{2014}]{webb2014}
{Webb} J.~J.,  {Leigh} N.,  {Sills} A.,  {Harris} W.~E.,   {Hurley} J.~R.,
  2014, \mn@doi [\mnras] {10.1093/mnras/stu961}, \href
  {http://adsabs.harvard.edu/abs/2014MNRAS.442.1569W} {442, 1569}

\bibitem[\protect\citeauthoryear{{Zocchi}, {Gieles}, {H{\'e}nault-Brunet}  \&
  {Varri}}{{Zocchi} et~al.}{2016}]{zocchi2016}
{Zocchi} A.,  {Gieles} M.,  {H{\'e}nault-Brunet} V.,   {Varri} A.~L.,  2016,
  \mnras, \href {http://adsabs.harvard.edu/abs/2016arXiv160502032Z} {in press}

\bibitem[\protect\citeauthoryear{{Zonoozi}, {K{\"u}pper}, {Baumgardt}, {Haghi},
  {Kroupa}  \& {Hilker}}{{Zonoozi} et~al.}{2011}]{zonoozi2011}
{Zonoozi} A.~H.,  {K{\"u}pper} A.~H.~W.,  {Baumgardt} H.,  {Haghi} H.,
  {Kroupa} P.,   {Hilker} M.,  2011, \mn@doi [\mnras]
  {10.1111/j.1365-2966.2010.17831.x}, \href
  {http://adsabs.harvard.edu/abs/2011MNRAS.411.1989Z} {411, 1989}

\bibitem[\protect\citeauthoryear{{Zonoozi}, {Haghi}, {K{\"u}pper}, {Baumgardt},
  {Frank}  \& {Kroupa}}{{Zonoozi} et~al.}{2014}]{zonoozi2014}
{Zonoozi} A.~H.,  {Haghi} H.,  {K{\"u}pper} A.~H.~W.,  {Baumgardt} H.,  {Frank}
  M.~J.,   {Kroupa} P.,  2014, \mn@doi [\mnras] {10.1093/mnras/stu526}, \href
  {http://adsabs.harvard.edu/abs/2014MNRAS.440.3172Z} {440, 3172}

\bibitem[\protect\citeauthoryear{{Zonoozi}, {Rabiee}, {Haghi}  \&
  {K{\"u}pper}}{{Zonoozi} et~al.}{2016}]{zonoozi2016}
{Zonoozi} A.~H.,  {Rabiee} M.,  {Haghi} H.,   {K{\"u}pper} A.~H.~W.,  2016,
  \mn@doi [\apj] {10.3847/0004-637X/818/1/58}, \href
  {http://adsabs.harvard.edu/abs/2016ApJ...818...58Z} {818, 58}

\bibitem[\protect\citeauthoryear{{Zotos}}{{Zotos}}{2015}]{zotos2015}
{Zotos} E.~E.,  2015, \mn@doi [\mnras] {10.1093/mnras/stu2129}, \href
  {http://adsabs.harvard.edu/abs/2015MNRAS.446..770Z} {446, 770}

\makeatother
\end{thebibliography}

\section*{Acknowledgements}

This research was supported in part by Lilly Endowment, Inc., through its support for the Indiana University Pervasive Technology Institute, and in part by the Indiana METACyt Initiative. The Indiana METACyt Initiative at IU is also supported in part by Lilly Endowment, Inc.  ALV is grateful to Douglas Heggie for stimulating conversations, and to the EU Horizon 2020 program for support in form of a Marie Sklodowska-Curie Fellowship (MSCA-IF-EF-RI 658088). We wish to thank the referee for a careful report.

\bsp	
\label{lastpage}

\end{document}